\documentclass[fleqn,twoside]{article}
\usepackage{espcrc2}
\usepackage{epsfig}

\title{Schr\"odinger functional at $N_{\rm f}=-2$\thanks
{Talk given by B. Gehrmann at Lattice 2001, Berlin}}

\author{Bernd Gehrmann, Stefan Kurth, Juri Rolf and Ulli Wolff
\address{Institut f\"ur Physik, Humboldt-Universit\"at zu Berlin,
Invalidenstr. 110, D-10115 Berlin, Germany}
}

\begin{document}

\begin{abstract}
We study the Schr\"odinger functional coupling for lattice Yang-Mills
theory coupled to an improved bosonic spinor field, which corresponds
to QCD with minus two light flavors. This theory serves as a less 
costly testcase than QCD for the scaling of the coupling.
\end{abstract}

\maketitle

\section{Introduction}

The non-perturbative computation of the strong coupling constant
$\alpha_s$ from hadronic scales to perturbative high energy scales is
one of the main interests of the ALPHA collaboration. For this
computation, a finite volume renormalization scheme is used in which
the coupling runs with the space-time volume \cite{Luscher:1991wu}.
The scale evolution of the coupling is computed recursively with the
help of the step scaling function, which is an integrated variant of
the beta function for finite changes of the scale. Furthermore, 
Schr\"odinger functional boundary conditions and ${\rm O}(a)$ 
improvement are employed. An important property of the step scaling 
function is that it can be calculated by simulating pairs of lattices 
with different sizes $L$ for the same $a$ and taking the continuum limit.

After this programme had been successfully used in the quenched
approximation \cite{heplat9309005}, first results for full QCD with
two flavors are now available \cite{Frezzotti:2000rp}.
Unfortunately, full QCD is notoriously costly to simulate, and thus
the data for the step scaling function do not reach very close to the
continuum limit. In order to learn more about the extrapolation to the
continuum and about the effect of ${\rm O}(a)$ improvement, we have
studied the step scaling function in a Yang-Mills theory coupled to a
bosonic spinor field, which has a local interaction and is therefore
much cheaper to simulate. It corresponds to setting the number of
flavors $N_{\rm f}=-2$ in the partition function, and is therefore closely
related to full QCD in perturbation theory. In the literature, this
model is known as \emph{bermion model} \cite{deDivitiis:1995yz}.
Our focus here however is not on extrapolating results from negative
to positive flavor number.

\section{Bermion model}

For a detailed discussion of the lattice setup, boundary conditions 
and the algorithm we have used, we refer to \cite{heplat0106025} and 
references therein.

The Schr\"odinger functional, as the partition function of
the system, is an integral over all gauge and quark fields which
fulfill the given boundary conditions. After integrating out the
quark fields, it is
\begin{eqnarray}
  Z = e^{-\Gamma} &=& \int D[U] D[\bar\psi] D[\psi]
                       e^{-S[U,\bar\psi,\psi]} \nonumber\\
                  &=& \int D[U] e^{-S_g} \det(M^\dagger M)^{N_{\rm f}/2}
\end{eqnarray}
with a gauge action $S_g$. For $N_{\rm f}=-2$, the determinant can be
written as
\begin{equation}
  Z = \int D[U] D[\phi^+] D[\phi] e^{-S_g-S_b},
\end{equation}
with a now local bosonic action
\begin{equation}
  S_b[U,\phi] = a^4 \sum_x |(M\phi)(x)|^2.
\end{equation}
The fields $\phi(x)$ carry color and Dirac indices.

We have chosen the bulk improvement coefficient $c_{\rm sw}$ by
extrapolating the non-perturbative data for $N_{\rm f}=0$ 
\cite{heplat9609035} and $N_{\rm f}=2$ \cite{heplat9803017} in 
$N_{\rm f}$. An explicit calculation of $c_{\rm sw}$ along the
lines of these references for the most critical parameters in
our simulations has proven a good accuracy of this extrapolation.
The boundary improvement coefficients $c_t$ and
$\tilde c_t$ have been set to their perturbative values.

\begin{figure}[htbp]
  \begin{center}
    \vspace{-1.5cm}
    \epsfig{file=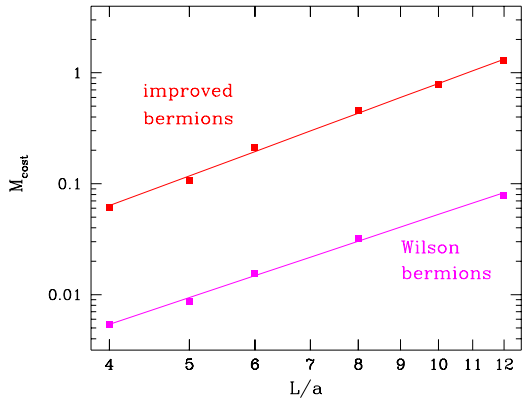,height=7.5cm,width=7.5cm}
    \vspace{-1.5cm}
    \caption{\label{fig:perf}
      \sl Cost for the measurement of $\sigma(0.9793)$ for Wilson
      and improved bermions.}
    \vspace{-1.0cm}
  \end{center}
\end{figure}

As the improved bosonic action depends quadratically on
each link $U_{x\mu}$ through the clover term, it is difficult
to employ finite step size algorithms such as the combination
of heatbath and overrelaxation steps used in \cite{heplat9907007}.
Because of the large additional cost of hybrid Monte Carlo
algorithms, we have decided to use global heatbath steps for
the bosonic fields and local overrelaxation steps with respect
to the unimproved action for the gauge fields. The clover term
is taken into account by an acceptance step. The acceptance rate
has turned out to be high enough (about 70~\%) for such an
approach. Nevertheless, there is a significant overhead due to
the calculation of the action difference necessary for the acceptance
step, resulting in a cost factor of about $12$ compared to 
unimproved Wilson bermions. On the other hand, comparing with
data from \cite{Frezzotti:2000rp}, improved bermions are about a factor 
of $10$ cheaper than dynamical fermions, with a better
scaling in $a$. In figure~\ref{fig:perf} we see a scaling of the
cost with $a^{-2.5}$.

\section{Results}

We have computed the step scaling function $\Sigma(u,a/L)$ for 
the couplings $u=\bar g^2=0.9793$ and $u=1.5145$ and lattice sizes
$L=4,5,6,8$. Most simulations were done on APE100 parallel computers
with up to 256 nodes with 50 MFlops each. For the simulation of
the largest lattice $L/a=16$ at $u=1.5145$ we have also used a
crate of APEmille.

\begin{figure}[hb]
  \begin{center}
    \vspace{-1.0cm}
    \epsfig{file=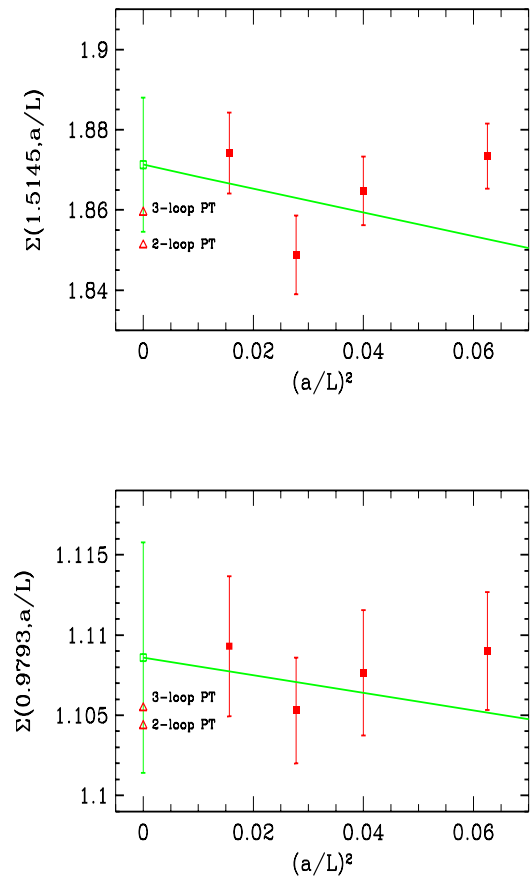,height=14cm,width=7.5cm}
    \vspace{-2.0cm}
    \caption{\label{fig:bermimp}
      \sl Step scaling function for improved bermions for the couplings 
      $u=0.9793$ and $u=1.5145$ with fits linear in $(a/L)^2$. Also shown
      is the extrapolated continuum value and the 2- and 3-loop
      results.}
  \end{center}
\end{figure}

Figure~\ref{fig:bermimp} shows the results for the step scaling
function plotted against $(a/L)^2$. Within the error bars, no 
linear dependence an $a/L$ is visible, and therefore an extrapolation
linear in $(a/L)^2$ is justified. The extrapolated values are consistent
with perturbation theory, and their error bars are of the same size
as the perturbative 3-loop contribution.

%
%

\begin{figure}[htbp]
  \begin{center}
    \epsfig{file=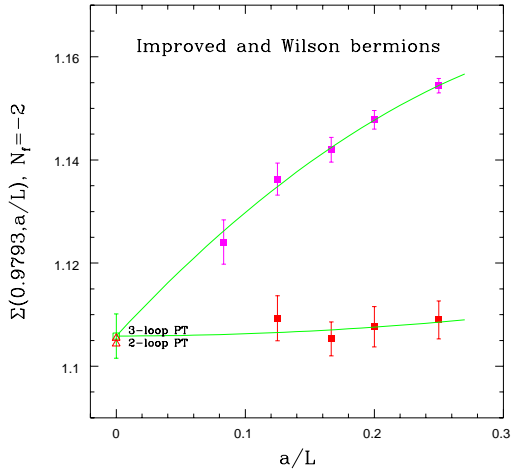,height=7.5cm,width=7.5cm}
    \vspace{-2.0cm}
    \caption{\label{fig:bermcombined}
      \sl Results for the step scaling function at $u=0.9793$, together
      with a quadratic fit under the constraint of universality.}
    \vspace{-1.0cm}
  \end{center}
\end{figure}

In~\cite{heplat9907007}, the step scaling function was already
computed in the unimproved bermion theory for $u=0.9793$.
Figure~\ref{fig:bermcombined} shows these results together with the
data after implementing improvement.  The linear
cutoff effects for this observable are of the order of a few percent.
In the plot, both data sets are fitted under the constraint of 
universality, i.e. that their continuum limit agrees.
This fit is linear plus quadratic in $a/L$ for the Wilson bermion data
and quadratic in $a/L$ in the improved data. Although the additional
input from the Wilson data is included, the joint continuum limit
$\sigma_{\rm combined}(0.9793)=1.1059(43)$ is almost the same as
the value $\sigma_{\rm improved}(0.9793)=1.1063(46)$. A linear 
plus quadratic fit in $a/L$ of the unimproved data alone would have 
given the continuum result $\sigma_{\rm unimproved}(0.9793) = 1.103(12)$.

This indicates a success of the improvement programme. The main
contribution to the cost for the calculation of $\sigma(u)$ comes
from the largest lattice. When using an improved action, the
lattice size needed for a reliable extrapolation to the continuum
is smaller than without improvement. Even in the model used here,
where the algorithmic implementation implies a large overhead for
the inclusion of the clover term, this leads to the improved case
being more cost effective. In simulations with dynamical fermions
and algorithms like Hybrid Monte Carlo, the implementation of
a clover term is possible with much lower overhead, and the 
advantage should therefore be higher.

\vspace{3mm}

{\bf Acknowledgements.}
We would like to thank Peter Weisz for essential checks
on our perturbative calculations and Rainer Sommer for 
helpful discussions.
DESY provided us with the necessary computing resources and the APE 
group contributed their permanent assistance.
This work is supported
by the Deutsche Forschungsgemeinschaft under 
Graduiertenkolleg GK 271 and by the
European Community's Human Potential Programme
under contract  HPRN-CT-2000-00145.

\end{document}